# Leveraging 'Social-Network' Infrastructure to Improve Peer-to-Peer Overlay Performance: Results from Orkut

Zahid Anwar[†‡]   William Yurcik[‡]   Vivek Pandey[†]   Asim Shankar[†]   Indranil Gupta[†]   Roy H. Campbell[†]

[†] Department of Computer Science
[‡] National Center for Supercomputing Applications (NCSA)
University of Illinois at Urbana-Champaign USA
{anwar,byurcik}@ncsa.uiuc.edu   {pandey1,shankar}@uiuc.edu   {indy,rhc}@cs.uiuc.edu

*Abstract*

*Application-level peer-to-peer (P2P) network overlays are an emerging paradigm that facilitates decentralization and flexibility in the scalable deployment of applications such as group communication, content delivery, and data sharing. However the construction of the overlay graph topology optimized for low latency, low link and node stress and lookup performance is still an open problem. We present a design of an overlay constructed on top of a social network and show that it gives a sizable improvement in lookups, average round-trip delay and scalability as opposed to other overlay topologies. We build our overlay on top of the topology of a popular real-world social network namely "Orkut". We show Orkut's suitability for our purposes by evaluating the clustering behavior of its graph structure and the socializing pattern of its members.*

## 1. Introduction

### 1.1. Background

Initially popularized by the file sharing applications such as Napster [1], Gnutella [2] and FreeNet [25], peer-to-peer has had an explosive growth in systems aimed at exploiting various aspects of resource sharing e.g. Digital Preservation Systems[34], Web Caching[30] and on-demand media distribution[26]. While much of the attention had been focused on the copyright issues raised by file sharing, peer-to-peer applications have recently been hailed by the research as well as industrial community because of their many attractive technical aspects like (1) decentralized control, (2) self-organization, (3) adaptability and (4) scalability. Peer-to-peer (P2P) systems can be characterized as distributed systems in which all nodes have identical capabilities and responsibilities and all communication is symmetric.

As opposed to the classical client-sever architecture with P2P, clients on a network can simply bypass the server and exchange information over the network directly. This adds value to the edges of a network where the information is being collected and used.

Gnutella – a popular file sharing application, broadcasts messages to all the peers in the path of the query. A querying peer sends the query to all of its peers, who in turn send the query to all of their peers until the query reaches a peer that produces a hit matching the query. This peer sends back a reply containing its address, the size of the file, speed of transfer, etc. The reply traverses the same path but in reverse order back to the querying peer. This passing of messages generates much traffic in the network, often leading to congestion and slow responses. Since responses to queries are delayed, quality-of-service becomes poor. We attempt to show that an alternative model where each node chooses its peers based on a metric of closeness (we use social interests) not only improves the overall search time but also leads to improved geographic proximity and hence lower communication delays.

This paper is organized as follows: In the remaining part of this introductory section we look at related research and outline the uniqueness of our approach. This is followed by Section 2 which provides a brief introduction to social networks. Section 3 describes how we analyzed the Orkut social network using a distributed Web Crawler. Section 5 analyzes the Orkut structure for hints on how it can be used to form an interest-based routing scheme. In Section 6 we describe a simulation of a P2P overlay based on Orkut. We end with a summary and conclusions in Section 7.

### 1.2. Related Work

We are not the first to study the use of social networks as the basis for forming a routing substrate for distributed hash tables. Work with the JADE multi-agent system platform [43] investigates the use of social networks to optimize the speed of search and improve the quality of service. Over a period of time nodes maintain a 'friends list' of peers in different interests groups that they see as being very similar to them and generate queries with high probability in those groups as compared to others.

Work on P2P Communities [31] use simulations of simple formation and discovery algorithms to show that structuring P2P networks according to interest-based communities can provide better search operations. The simple but powerful observation that using the same intuitive method of search human use - asking "those in the know" – improves both search quality and messaging. This is a similar idea to what we want to accomplish but this work created their own topology for P2P by crawling websites and assumed web sites are analogous to peers. Furthermore, they introduced certain rules to enforce a social network behavior when they discovered that their graph's path length and clustering coefficients actually did not exhibit small world behavior.

Two projects deployed in university settings known as "Buddy Web"[44] and "Best Peer"[37] try to improve search by employing a 'routing by similarity of interests' scheme. However these systems are limited to small-scale internal networks only and do not scale to the Internet.

To the best of our knowledge, this is the first paper to evaluate the behavior of a P2P system that uses an Internet-based social network consisting of millions of users as its underlying infrastructure. The experimental results we document have design implications for both the scalability and performance of future group communication systems.

## 2. Social Networks

### 2.1. Six Degrees of Separation, Small Worlds, and Scale-Free Networks

The "Six Degrees of Separation" phenomenon was first investigated by Stanley Milgram [36] in 1960 where he addressed letters to a particular stockbroker in New York and gave them to people randomly picked at locations in the United States far away from that of the final receiver. The condition for passing the letter, so that it reaches the addressee, was that one could post it only to people they knew personally by first name. Eventually most of the letters reached the destination and the average number of hops was six. Since then there have been various studies demonstrating how this effect may help people conduct their everyday lives.

Studies on information propagation have shown that unlike *strong ties* (within communities), *weak ties* help knowledge propagation across communities [29]. Social networks can be studied by computer simulation to investigate the evolution of societies of artificial agents simulating real people. Simulations allow us to study patterns, diversity and behavioral changes in groups of people due to changes in environment [40]. Social networks can also be studied using visualization techniques to show how strong are the relations between people in a group and between groups of people [28].

This effect, also known as "Small Worlds" or "Scale-Free Networks", has been revisited with analytical techniques starting with the seminal work by Barabasi [16, 17, 18]. Barabasi studied many natural and man-made networks and found that they all exhibit degrees of clustering with hub and spoke topologies and remote links betweens clusters. These real networks are fundamentally defined by a few highly connected nodes but even a very small number of remote links (weak ties) are sufficient to dramatically decrease the average separation between nodes.

Analytically Barabasi measured this clustering effect with power-law distributions showing varying power law exponents[1] for networks such as movies (by their actors especially Kevin Bacon), members of an audience (through auditory cues), social systems (family ties, school ties, friendships, etc.), biological organisms (biochemical signals), the brain (neural interconnections), and especially the Internet. In fact there have been a series of studies of the structure and topology of Internet-based networks best summarized in [39] including the www, Email, instant messaging, virus/worm propagation, and P2P networks. Before this work identifying and quantifying the scale-free nature of the Internet, every new algorithm proposed by researchers for improving network performance was typically tested on random networks generated by consensus tools (such as the Waxman Network Topology Generator) which in retrospect resulted in incorrect solutions which should now be reexamined.

### 2.2. The Orkut Community- A Social Networking Service

Orkut [3] is a social networking service based on the idea similar to Friendster [4] the popular online-relationships service that cleverly assimilates real-life social groups into a large virtual network and others like Everyone's Connected [5], Ryze [6], Ecademy [7] and LinkedIn [8]. These tools form a virtual community designed to help users meet new friends and maintain existing relationships. Orkut was quietly launched on January 22, 2004 by Google. The service was created by Google employee Orkut Büyükkökten, who had developed a similar system, "InCircle," for his previous employer, Affinity Engines. InCircle was intended for use by university alumni groups. Unlike the rest of the services, Orkut expanded to over millions of users at a remarkably fast pace.

---
[1] Power law exponents are referred to as "clustering coefficients" in the Small World terminology

## 2.3. Statistics

The time at which this study was conducted (late 2004) Orkut had a little over two million users. By the time the paper was completed the number of users had grown to over three million. We did not study all three million users but only about one-third. Since we began crawling the network from over 500 different random starting nodes we believe that our analysis is a valid approximation to the actual network. An interesting fact about the network is the male to female ratio of participants as shown in Figure 1.

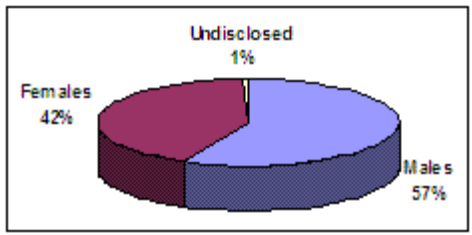

**Figure 1. Gender Distribution in Orkut**

## 2.4. Membership and Policies

A user can become a member of Orkut by invitation only but once a member he has the privilege to browse anyone's profile if he is connected to him via some friend or a friend of friend and so on. Users can choose to make available information such as age, hobbies, interests, movies, sports, region to everyone or to a selected circle of friends. In this study we aggregated only user information open to public view and did not keep a track on any user specific information in accordance with the community's privacy policies.

## 3. Distributed WebCrawler Design

The architecture of our distributed web crawler is shown at the top of Figure 2. We wanted to have a 'snapshot' view of the Orkut graph but the trouble with conventional crawlers is that by the time they join the different pieces of the puzzle to make the complete picture, the picture will be altered too much to make any sense. Simply put, we want the crawler to be fast enough so that the dynamic system we are exploring does not change enough to make our results invalid.

Figure 3 shows the meta-data information of our tuple space. Each node in the Orkut network is identified by a unique id. Producer nodes populate the space tuples representing partitions of a problem. We used the model of Yale Linda Group's Linda Coordination Language [20]. The Linda model defines a shared "tuple-space," which functions like a blackboard in which nodes in a distributed system can read-from and write-to.

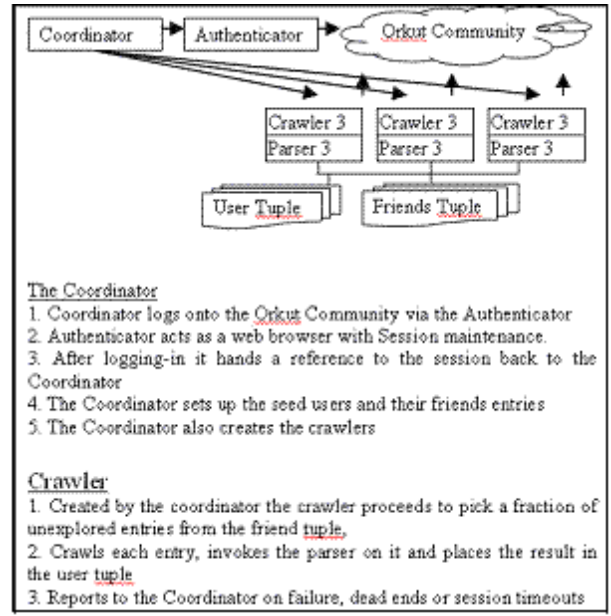

**Figure 2. Design of Our Distributed Web Crawler**

The bottom of Figure 2 shows the algorithm for our distributed crawler which is comprised of five main parts: (1) the coordinator, (2) authenticator, (3) crawler, (4) parser and (4) the distributed data structure for Tuples which can be accessed simultaneously by multiple crawlers. A simplified algorithm for how our crawlers (we use several crawler instantiations in parallel) explore the entire Orkut graph is as follows:

- select a node randomly from the graph and mark nodes successively using the standard breadth first search algorithm
- when crawlers reach nodes which have been discovered previously, they are stopped
- ultimately all the crawlers stop

| ID | Name | Friends Count | Age | Sex |
|---|---|---|---|---|
| Passions | TV Shows | Movies | Music | Books |
| Sports | Activities | Hometown | Country | Region |

**Figure 3. Meta Data Schema for Each Orkut User**

The system was implemented in Java and uses the HttpUnit package for parsing dynamic html content. The Web Crawler ran uninterrupted from Nov $28^{th}$ to Nov $30^{th}$ – a total of three days and gathered about 22 Gigabytes of data.

## 4. The Orkut Graph Structure

Before we attempt to leverage the graph structure of Orkut for performance we were interested in answering

the following questions which have direct impact on our results:

- How are the users distributed around the world?
- Are users evenly distributed or clustered in specific regions?
- What is the average frequency of the outgoing edges per node?
- How are different continents connected together?

### 4.1. Geographical Visualization of a Social Network

We used the country/hometown/region fields in the user tuple shown earlier to construct a visualization of the Orkut network. We used CAIDA's GeoPlot tool [10] to draw Orkut users on a map of the world. The country/regions were mapped to latitude/longitude coordinates by posting each user to "MapQuest"[11] and "How Far Is It?"[12] online web services. The distribution obtained was an approximation because less than 70% of the users choose to disclose their location and only 38% of these users actually pinpointed their city. As users do not select their locations from a predefined list, a significant minority of cities are misspelled or displayed in fancy nicknames which often confuses our parser. Nevertheless we achieved a reasonable bound on the node densities in various regions of the world.

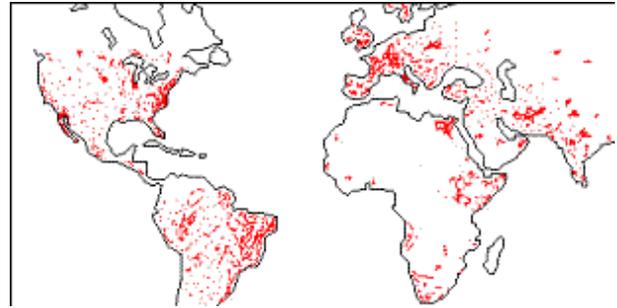

**Figure 4. Geographical Distribution of Orkut Users**

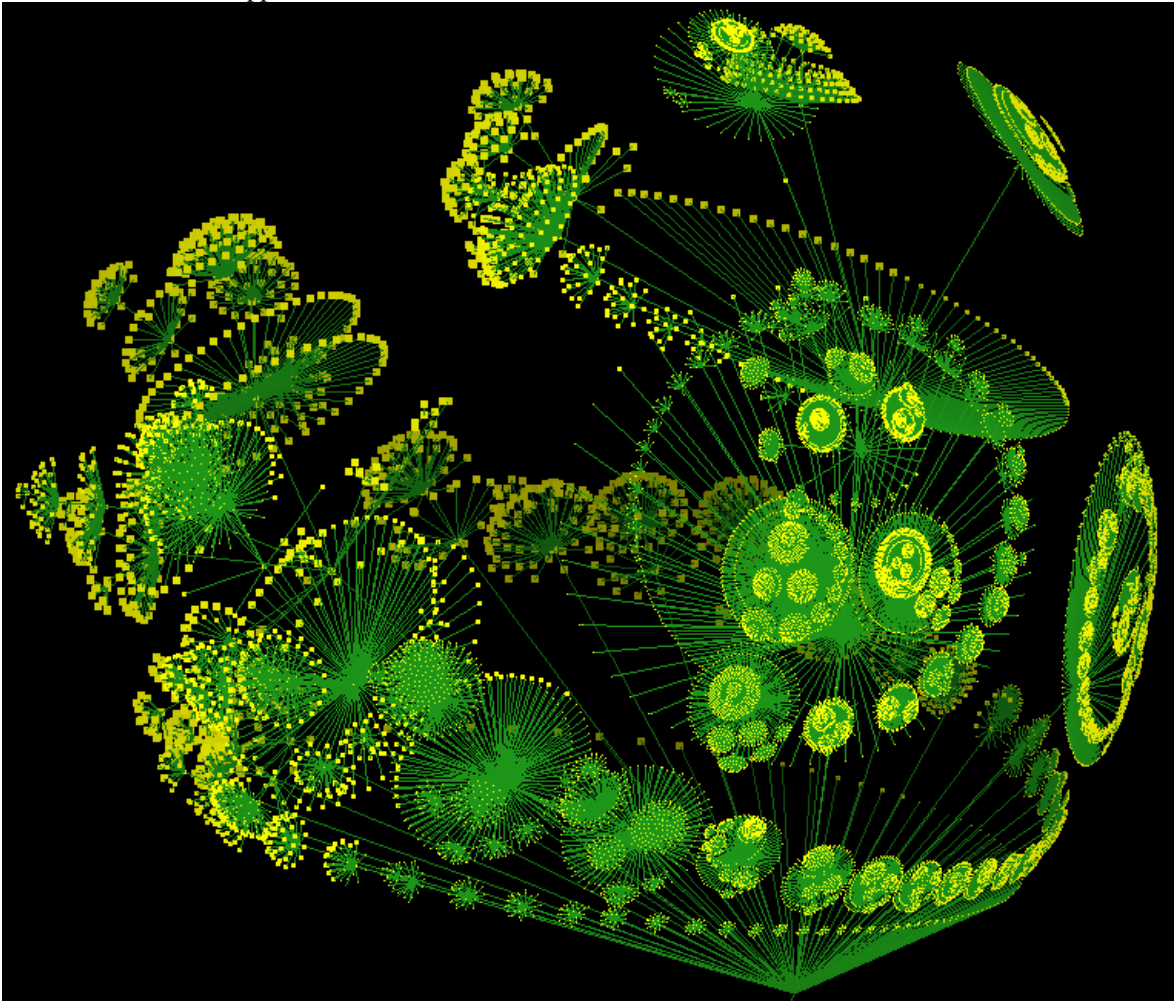

**Figure 5. The Orkut Graph Structure Visualized in Walrus**

Not unexpectedly we found the graph depicting the characteristics of a 'small world'[42] like a coauthorship network or the power grid of the western United States and Canada, with short paths between any two nodes and very high clustering. It also portrayed large degree of 'transitivity' meaning that if A is linked to B and B is linked to C, there is a high probability that A will also be linked to C.

Figure 4 shows the geographic distribution of Orkut nodes on the globe. A large percentage (53%) of Orkut's members were from Brazil, followed by 18% from the United States and 8% from Iran. It was also found to be popular in eastern Europe, south and south east Asia while only a small percentage in China and Russia.

Figure 5 uses the Walrus 0.6.3 [9] and Libsea 1.0 utilities for visualization of the Orkut graph structure. Walrus computes its layout based on a user-supplied spanning tree.  It is possible to specify multiple spanning trees. In the figure a spanning tree with 62911 links was specified. This characteristic tree was harvested from a subset of the original data using the Kruskal Algorithm to construct a spanning tree. It is interesting to note friends are grouped in 'clusters' with some nodes with high connectivity serving as hubs connecting many local nodes. The purpose of this visualization is to show how majority of the links are short (e.g. friends in the same city or country) and how a small subset of nodes (marked with large dots) act as hubs to the rest of the world by maintaining a huge network of links.

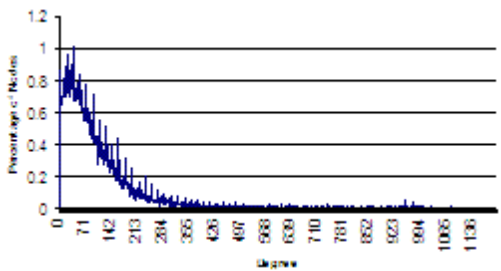

**Figure 6. Node Degree in the Orkut Network**

### 4.2. Visualizing Node Linkage

We also observe the node degree in the Orkut network, (i.e. the number of outgoing links to immediate friends). It can be observed from Figure 6 that the distribution resembles a power-law with a heavy tail similar to the Zipf distribution. However an absence of a straight line in the corresponding log-log plot showed that it is not completely power law. The maximum node degree we found was 1077 of a person in Argentina. This behavior is an outlier, however, the peak at a node degree of 19 shown in Figure 6 is the average connectivity.

## 5. Interest-based Routing

### 5.1. Popular Peer-to-Peer Routing Techniques

The subject of how to do routing in overlay networks is an active area of research. There are a variety of routing schemes, for instance Scribe[21], Split Stream[22] and Bayeux[46] that use a prefix based routing mechanism in a tree based fashion. The CAN[41] overlay uses a hypercube, Bullet[32] uses a Mesh while Scattercast[23] uses a hybrid approach.  These schemes can be thought of as static routing techniques optimized for various goals such as having at most log(n) lookup time and reducing link stress. There are also more dynamic gossip-based routing schemes [19,27 ,33] that primarily address the high presence of churn in P2P systems. Interest-based routing falls in between these two extremes in that it is more dynamic than static routing techniques and does not cope well with membership changes. Although it does not provide an upper bound to routing hops such as in prefix-based routing, it does provide more control to the user and typically reduces the hop count.

### 5.2. Applicability of Interest based routing

Interest-based routing does not target any special application domain but suggests a plethora of potential everyday applications such as file sharing, resource sharing, web caching, media distribution and group communications. These category of applications fall nicely into the social networks paradigm because they can take advantage of the knowledge of peers with similar interests. For example a friend who is the same nationality, age group and plays the same sports is very likely to also browse the same News websites and go to the same online stores as his peer. This corporate web caching scenario where each request does not have to go to a server but instead can be replied to by a peer who fetched the same page recently. This also has advantages when dealing with flash crowds [38].

### 5.3. Evaluating Strength of Orkut Interest Groups

How far do Orkut friends share interests? Since we were interested in constructing a Distributed Hash Table on top of this social network we examined how a group of friends are similar in their interests. We picked at random a group of 1000 people in the Orkut community and for each user began a breadth-first search starting at their immediate circles of friends (hop 1), friends of friends (hop 2) and so on looking for a match in interests. So for example if a person says that he is interested in 'Science Fiction' movies then we looked for another 'SciFi' fan in his tree.

Our search latency results are shown in Figure 7. Interestingly we found that even close friends can have a very different taste in movies and TV shows however the same does not hold for music and sports. So even though the probability of finding a fellow 'movie or TV show' fan goes up exponentially depending on the depth of the level being searched, the probability of finding someone with the same taste in music or sports is even higher. On average it took three hops only to find similar interests (with a probability of 97%). With parallel search time in the same level of children and an average node degree of 19 this result can be leveraged for performance optimization in an overlay.

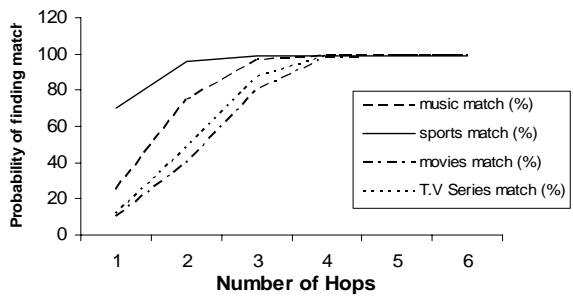

**Figure 7. Average Latency of Search Query**

## 6. Overlay Simulation

The idea is to utilize this centralized information stored at the Orkut servers to make a distributed overlay. We do not consider an overlay formation algorithm in this paper but assume that one or more Orkut servers provide the joining node the location information (e.g. IP address) to connect to his peers. Once a node has registered and connected with all its peers then it can start sending queries and receiving responses. When a peer finds that it has the reply to a query that was routed to it then it replies back directly to the requester. The requester can then mark that peer as a potential server for that particular categories of requests. Along with the entry of that server, the peer can put a weighted strength for its relationship with the server depending on experience of previous query replies from the server and the precision of the current search result. This way the routing tables of the peers evolve over time so that even if the requesting peer was unlucky to have poorly matched friends in the beginning, they can still improve over time.

Note that in the real Orkut network, if you belong in someone's friend list then they will also belong in yours. We can mimic this behavior by having the responding server add an entry of the requester to his friends list (after all if both peers like the same file then they might have other resources in common as well). Nodes can choose to throw away peers that hardly ever respond to their queries.

### 6.1. Simulation Description

We simulates a real scenario to prove our earlier assertion that a peer-to-peer network built using the Orkut topology greatly optimizes the search query time. A secondary motive was to see if an underlying social network infrastructure leads to lower delays in group communication by fetching information from peers that are geographically close. For this purpose we evaluated the performance of the data delivery mechanism in terms of average delay and group size.

The simulation environment is a transit-stub topology of 5,000 nodes generated using the Georgia Tech random graph generator [45] using the transit-stub model. We used this model because it accurately reflects the wide range of properties of real internetworks, including hierarchy and locality. We then mapped a small subset of the Orkut network onto this topology. Friends with the smallest distances between them (also calculated using the "How Far Is It?" website) were placed on the same stub domains whereas the ones further apart were placed across transit domains. There were 20 stub domains per transit node, with an average of 20 routers in each. There were 10 transit domains with each (on average) having 25 nodes, and an edge between each pair of nodes with probability 0.6. Each stub domain had (on average) eight nodes, and edge probability 0.42.

### 6.2. Simulation Results

We used the myns simulator[13] implementation of the ESM[24] and NICE[14] protocols topology for comparison with the Orkut topology. ESM which is based on the Narada Overlay constructs its trees in a two-step process by first constructing a rich mesh with heuristics for achieving shortest path delays between any pair of members and a limited number of members per node. Thereafter it constructs a reverse shortest path spanning tree from the mesh. The NICE protocol arranges the set of end hosts into a hierarchy; the basic operation of the protocol is to create and maintain the hierarchy. The hierarchy implicitly defines the multicast overlay data paths. The members at the very top of the hierarchy maintain (soft) state about O(log N) other members. While constructing the NICE hierarchy, members that are *close* with respect to the distance metric are mapped to the same part of the hierarchy allowing the protocol to produce trees with low stretch.

Different group sizes of friends were chosen from the Orkut network for measuring multicast delivery delays. For each group of nodes we also ran the ESM and NICE multicast algorithms and found out that that forwarding data along Orkut links to your friends generally gives lower delays than over a topology generated by ESM and NICE (see Figure 8). Our assumption here was that only groups of friends want to multicast information to each other and the results might be different if we had chosen a random group of multicast peers.

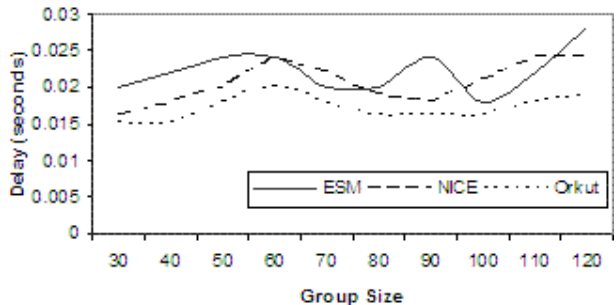

**Figure 8. Average Multicast Data Delivery Latency**

The Orkut routing metric was to forward packets on to neighbors with the highest degree. Such high-degree nodes (use a minimum threshold for degree to decide such *supernodes* and maintain further information about distances to other nodes in the vicinity. We plan to do a comparison with the standard *distance vector* routing techniques in a next paper.

ESM showed a large fluctuation of delay as the group sizes increased. We believe this is because of our assumption that only groups of friends want to multicast information to each other and the results might be different if we had chosen a random group of multicast peers. It may also be that the group size we used was small (a maximum of 130 members) where as experiments with delays in ESM [24] involved very large sizes and showed little fluctuation. The NICE protocol showed more steady performance but the hierarchical overlay it constructed had more delay than Orkut. In [15] the authors mention that they could incorporate topology-awareness in the algorithm to achieve better performance but the version we used for our simulations did not have that modification.

In the future we would like to evaluate our system using larger group sizes and incorporate features such as self-formation, survivability, security and adaptability around failures.

## 7. Conclusions

In this paper we evaluated a real world social network consisting of over 3 million users. We observed social patterns in the system and saw how closely matched friends are in their interests. We next examined how the topology lends itself to being used for setting up links in a Distributed Hash Table scenario. We then simulated a distributed system based on the Orkut topology and showed that not only is it reduce search latency but that it also reduces delays in group communication compared to other application-level multicast techniques (specifically ESM and NICE).

Although it is not clear how users will embrace open peer-to-peer social networks in the long term [35], based on Internet-scale simulations we have shown that overlays leveraging social network infrastructure can reduce delays and scale to large sizes.